\documentclass{svjour3}
\smartqed  
\usepackage{graphicx,amssymb}
\journalname{Experimental Astronomy}

\begin{document}

\title{Simulations of Galactic Cosmic Ray Impacts on the Herschel/PACS bolometer Arrays with Geant4 Code}

\author{C.Bongardo \and P.M.Andreani \and K.Okumura \and B.Horeau \and A.Claret \and G.De Zotti \and R.Giannitrapani \and F.Longo}

\institute{C. Bongardo \at
              INAF--Osservatorio Astronomico di Padova, Vicolo dell`Osservatorio 5, I-35122 Padova, Italy \\
              \email{christian.bongardo@oapd.inaf.it} \and
              P.M. Andreani \at
              European Southern Observatory, Karl Schwarzschild strasse 2, D-85748 Garching near Munich, Germany\\
              \email{pandreani@eso.org}\\
              \emph{on leave from: INAF--Osservatorio Astronomico di Trieste, via Tiepolo 11, I-34143 Trieste, Italy} \and
              K. Okumura \at
              DSM/DAPNIA/SAp CEA--Saclay, Bat. 709, l`Orme des Merisiers, F-91191 Gif-sur-Yvette CEDEX,
              France \\
              \email{koryo.okumura@cea.fr} \and
              B. Horeau \at
              DSM/DAPNIA/SAp CEA--Saclay, Bat. 709, l`Orme des Merisiers, F-91191 Gif-sur-Yvette CEDEX,
              France \\
              \email{benoit.horeau@cea.fr} \and
              A. Claret \at
              DSM/DAPNIA/SAp CEA--Saclay, Bat. 709, l`Orme des Merisiers, F-91191 Gif-sur-Yvette CEDEX,
              France \\
              \email{arnaud.claret@cea.fr} \and
              G. De Zotti \at
              INAF--Osservatorio Astronomico di Padova, Vicolo dell`Osservatorio 5, I-35122 Padova, Italy \\
              \email{gianfranco.dezotti@oapd.inaf.it} \and
              R. Giannitrapani \at
              Universit\`a degli Studi di Udine, Via delle Scienze 208, I-33100 Udine,
              Italy \\
              \email{riccardo@fisica.uniud.it} \and
              F. Longo \at
              Dipartimento di Fisica, Universit\`a degli Studi di Trieste e INFN sezione di Trieste, Via Valerio 2, I-34127 Trieste, Italy \\
              \email{Francesco.Longo@ts.infn.it}}

\date{Received: xx/yy/200z / Accepted: xx/yy/200z}
\maketitle

\begin{abstract}
The effects of the in-flight behaviour of the bolometer arrays of the Herschel/PACS instrument under impacts of Galactic
cosmic rays are explored. This instrument is part of the ESA-Herschel payload, which will be launched at the end of 2008 and will
operate at the Lagrangian L2 point of the Sun-Earth system.
We find that the components external to the detectors (the spacecraft, the cryostat, the PACS box, collectively referred to
as the `shield') are the major source of secondary events affecting the detector behaviour. The impacts deposit energy on the
bolometer chips and influence the behaviour of nearby pixels. 25\% of hits affect the adjacent pixels. The energy
deposited raises the bolometer temperature by a factor ranging from 1 to 6 percent of the nominal value.
We discuss the effects on the observations and compare simulations with laboratory tests.

\end{abstract}

\section{Introduction}
The European Space Agency Herschel satellite will be launched at the end of 2008  with an Ariane-5 rocket and will operate at the
Lagrangian L2 point of the Sun-Earth system (see ESA web page: {\it www.rssd.esa.int/SA-general/Projects/Herschel}). Herschel
is the ESA fourth cornerstone mission and will perform imaging photometry and spectroscopy in the far infrared and
submillimetre part of the spectrum. The Herschel payload consists of two cameras/medium resolution spectrometers (PACS and
SPIRE) and a very  high resolution heterodyne spectrometer (HIFI).

The PACS (Photo-conductor Array Camera and Spectrometer) instrument will perform efficient imaging and photometry in three
wavelength bands in the range 60--210$\mu$m, and spectroscopy and spectroscopic mapping with spectral resolution between 1000
and 2000 over the same wavelength range or short segments.

PACS is made of four sets of detectors: two Ge:Ga photoconductor arrays for the spectrometer part and two Si-bolometers for
the photometer part. On each instrument side each detector covers roughly half of the PACS bandwidth\footnote{ More about
PACS can be found in the PACS Web pages: {\it pacs.mpe-garching.mpg.de} and {\it pacs.ster.kuleuven.ac.be}}.

It is well known that cosmic rays may influence strongly the detector behaviour in space. The performances of Infrared Space
Observatory (ISO) detectors were largely affected for time periods long enough to corrupt a large amount
of data \cite{Heras}. Our goal is to exploit our present knowledge about the detectors and the cosmic
environment to understand how the detector behaviour changes and how we may retrieve the lost
information and/or extract the astronomical signal from the sources through an appropriate data analysis
tool.

In this paper we focus our attention on the Si-bolometer arrays of the PACS instrument. These detectors differ from those on
board of ISO as they are not of photoconducting type and very likely less affected by cosmic ray hitting. In a previous paper
we dealt with the Ge:Ga photoconductors of the same instrument (\cite{Bon06}).

The paper is organized as follows: in Sections \ref{simulations} through \ref{PL}, we briefly describe the used simulation
toolkit, the input detector design, the Galactic cosmic ray spectra and the physics list. In section \ref{checks} we discuss very
simple tests to check the reliability of the present simulations. The results are reported in section \ref{results} and discussed
in section \ref{discussion}, while
in section \ref{IPN} we compare simulations with laboratory tests. Conclusions about the effect on observations are
to be found in \ref{Conclusion}.

\section{The Geant4 Monte Carlo code}\label{simulations}
Monte Carlo simulations were carried out with the Geant4\footnote{Every simulation was run with the 7.0 patch 01 version,
vith CLHEP 1.8.1.0 with a gcc 3.2.3 compiler} (hereafter G4) toolkit, which simulates the passage of particles through
matter. G4 provides a complete set of tools for all the domains of detector simulation: Geometry, Tracking, Detector
Response, Run, Event and Track management, Visualization and User Interface. An abundant set of physics processes handle the
diverse interactions of particles with matter across a wide energy range, as required by the G4 multi-disciplinary nature;
for many physics processes a choice of different models is available. For any further information see the Geant4 Home page
at: {\it geant4.web.cern.ch/geant4/}.

\section{The Detector Design}\label{volumes}
The PACS camera is made of two bolometers: one called `blue' working at short wavelengths (60-130$\mu$m) and one `red' at
long wavelengths (140-200$\mu$m). The blue one consists of 8 chips ($4\times 2$), each of dimensions of 15.78$\times$12.63 mm
and comprising $16\times 16$ pixels. The chips are separated from each other by 120 $\mu$m and have a projected field of view
in the sky of 1.75$^{\prime} \times$3.5$^{\prime}$. The red one is made of 2 chips covering the same sky area, each of
dimensions of 12.63$\times$12.63 mm and comprising $16\times 16$ pixels. 

The detector module of the PACS bolometers is made of different parts: each pixel consists of an absorber supported by a
silicon grid 5$\mu$m thick with a thermometer, and is suspended in an Inter-Pixel-Wall (IPW) grid 450$\mu$m thick which has a
reference thermometer and acts as a thermostat, of aluminum, Titanium and Silicon layers, of a Silicon strip, and of
reflecting slabs \cite{bolocea}. The output signal is a function of the temperature difference between the pixels and the
reference thermometers.

\begin{figure}
\centering
\includegraphics[height=4cm, width=8cm]{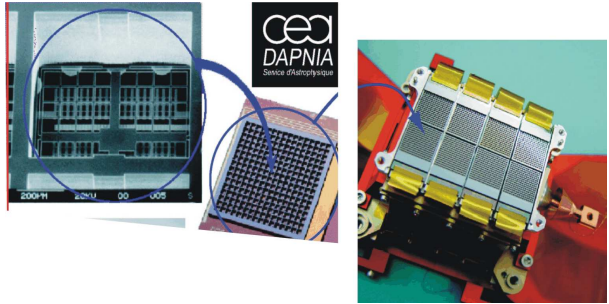}
\caption[The real PACS blue bolometer]{ Left: a highly magnified picture of one pixel (the Si grid) of size 0.75mm.
Such pixels are assembled into 16x16 pixel subarrays (middle), which in turn are assembled into the focal plane
matrices. Right: a picture of the focal plane PACS blue bolometer camera.
}\label{bluereal}
\end{figure}

\begin{figure*}
\centerline{\includegraphics[height=12cm, width=12cm]{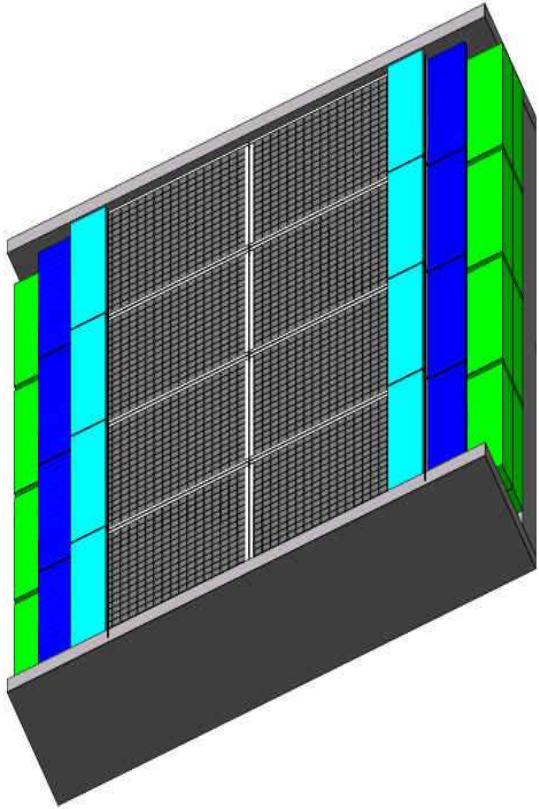}}
\caption[]{The PACS blue bolometer array. Grey: the Al layers. Green: the Ti layers. Blue: the Si layers. Cyan: the Si strip.
Black: the IPW.} \label{blue}
\end{figure*}

Figure \ref{bluereal} shows a picture of the real PACS blue bolometer detectors, while Figure \ref{blue} shows how we have
modeled their geometry. This last configuration is that used for for the simulations with the G4 software.
Details of the modeled geometry are given in Appendix A.

The ensemble of all the components `external' to the detector, i.e. the PACS box, the cryostat and the spacecraft is
hereafter called `shield'. The dimensions of this latter have been chosen to be big enough to contain the PACS detectors.
A detailed simulation should in principle include a model of the exact geometry of the spacecraft. For the PACS bolometer,
however, we approximate the spacecraft, the cryostat and the telescope with an aluminum sphere, 11 mm thick and at a
temperature of 80 K, which is the nominal temperature of the passively cooled Herschel telescope.
The inner radius of the 'shield' is 20 cm and the outer radius 21.1 cm. This aluminum sphere corresponds to 15.7 Kg, i.e.
5 percent of the telescope mass. We have carried out the same tests
described in this paper by varying the telescope temperature from the 80 K value and it turns out that the results do
not change.

The G4 software then requires the design of a {\it world volume} containing all the detector parts and the specification of the
environment in which the detector is placed. We made the {\it world volume} a little bit larger than the 'shield', i.e. a sphere
with a 21.2 cm radius.

\section{Galactic Cosmic Rays spectra}
Herschel will be injected in a orbit around L2. The Galactic Cosmic Ray (hereafter GCR) rate in the L2 environment has been
measured by the WMAP satellite \cite{Prantzos} and found to be very close to that of the geostationary orbit, for which the
GCR spectra are known. We use the Cosmic Ray Effects on Micro-Electronics (CREME) model set both for the solar-minimum (GCRs
maximum) and for the solar-maximum environment \cite{CREME}. CREME treats geosynchronous and near-Earth interplanetary
environments as identical, there is no geomagnetic shielding and no trapped particles \footnote{Strictly speaking, this is
not correct, but the differences arise at energies below $\sim$ 5 MeV/nuc, which are irrelevant for single-event effects.}.
The CREME interplanetary flux model (galactic cosmic rays, anomalous cosmic rays, and solar energetic particles) is based
on measurements at Earth (1 AU). CREME uses numerical models for the ionizing radiation environment in near-Earth orbits to
evaluate the resulting radiation effects on electronic systems in space\footnote{In this work we used the CREME86 version,
which was more extensively tested and compared with measurements. An updated version, CREME96, is available at: {\it
https://creme96.nrl.navy.mil/}}.

\begin{figure*}
\centerline{
\includegraphics[height=6cm, width=6cm]{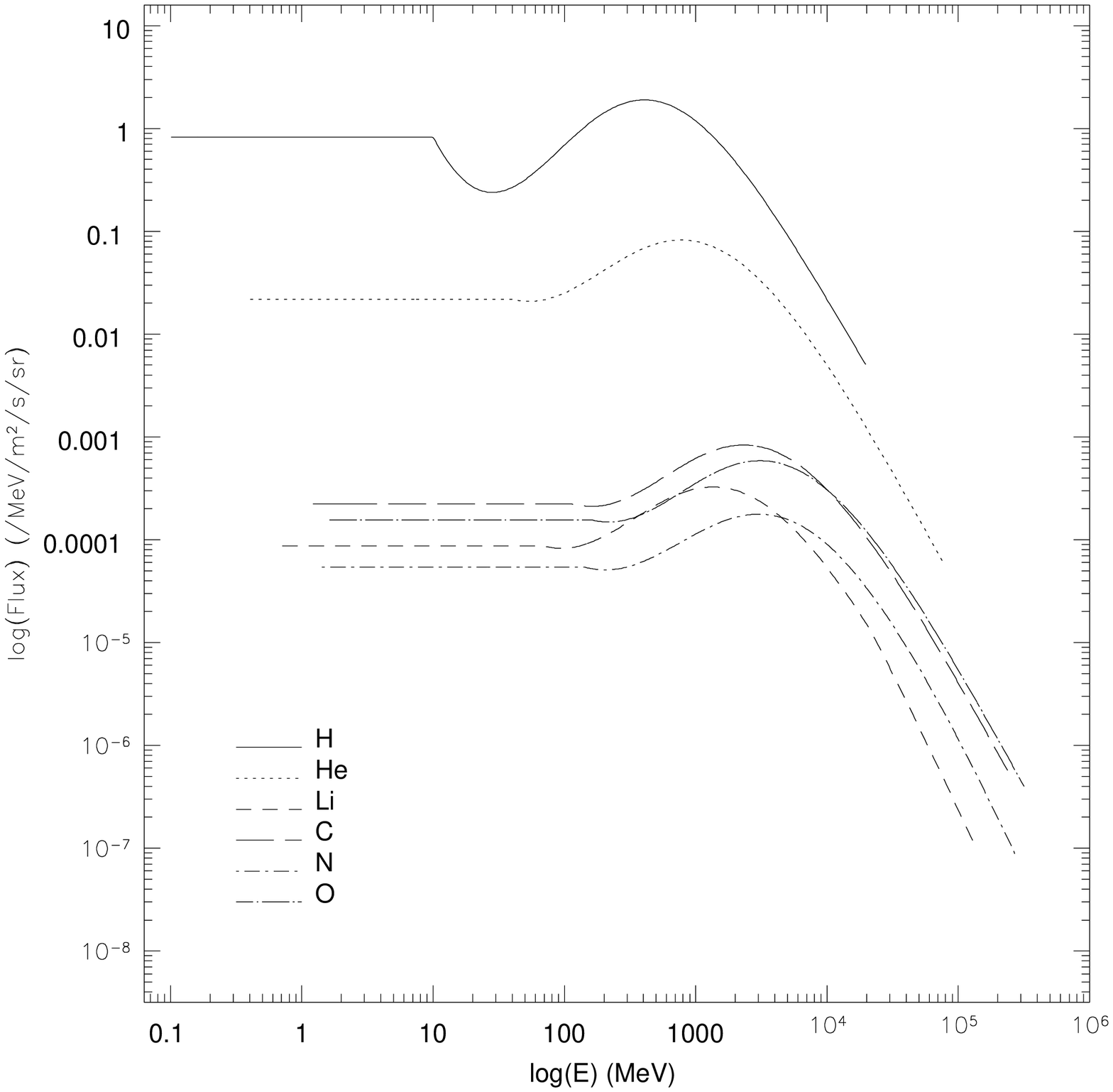}\\
\includegraphics[height=6cm, width=6cm]{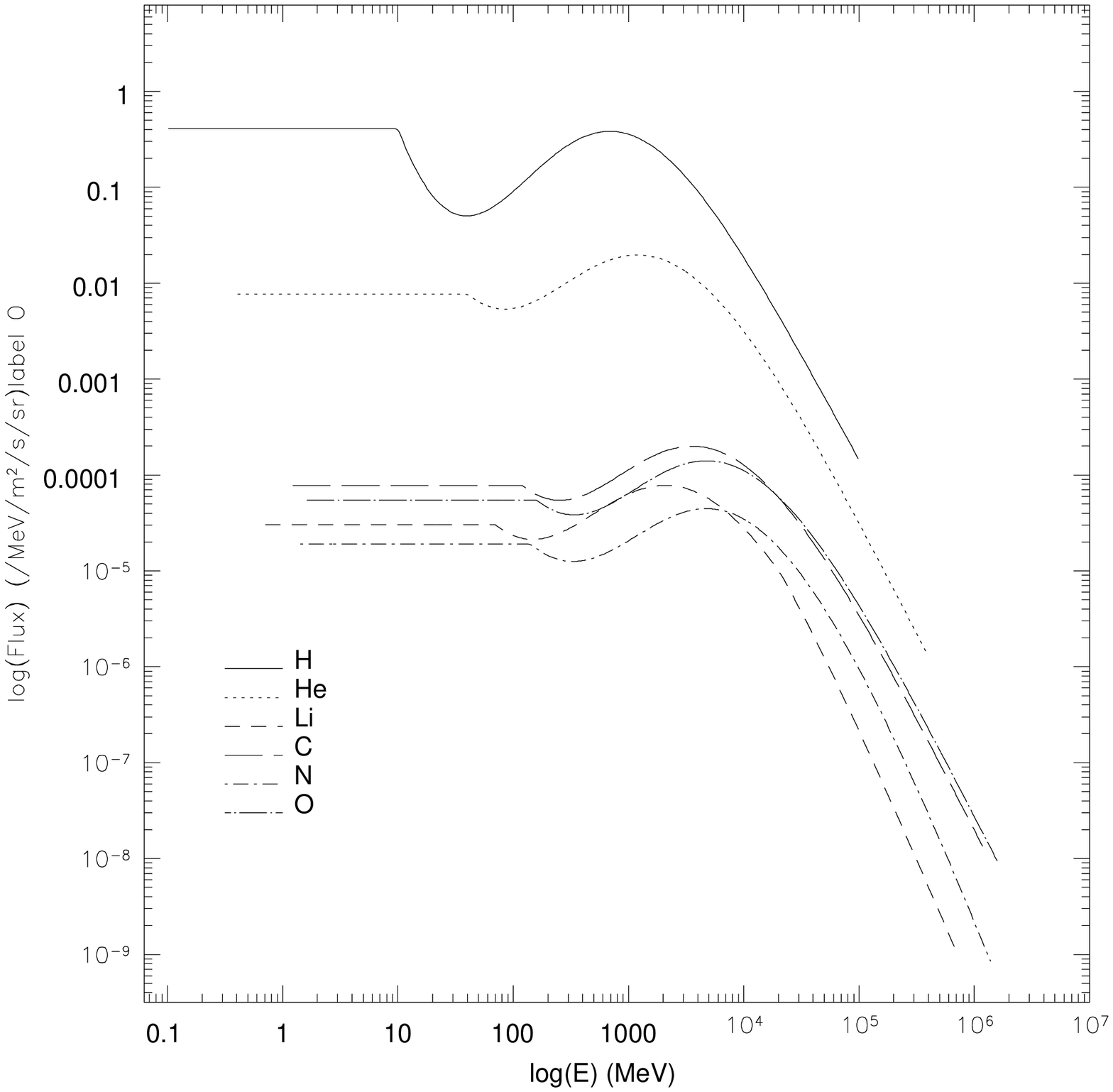}}
\caption[]{\small Differential GCR Spectral Energy Distributions (SEDs)
at the solar minimum (left) and maximum (right). Each curve corresponds to a type of
particle; from top to bottom: protons, $\alpha$, C, O, Li, N. Values are from CREME86. 
} \label{GCR_SED}
\end{figure*}

Our simulations take into account the most common GCR particles in the near-Earth heliosphere: protons, alpha particles and
nuclei of Li, C, N, O (see Figure \ref{GCR_SED}). We had to neglect heavier ions as physical processes in Geant4 lack a detailed
comparison with experimental data and reliability. Fe ion, although with an abundance lower of two order of magnitudes than
that of Oxygen may cause a big jump in the detector temperature because of its larger mass. The current version of G4 is
not reliable to follow these processes and they are then neglected in this work.
In Table \ref{GCR_rates} we report the rates (number of particles per
second) of each GCR type considered, computed by integrating the fluxes shown in Figure \ref{GCR_SED} over the 'shield'
surface and particle energy. We have considered here that the detector is hit by the flux coming from all directions
and that a Herschel will reach the L2 point at a solar minimum but it will operate until an epoch
of increasing solar activity. Both phases of solar activities are thus considered.

\begin{table}
\caption[]{Rates (number of nuclei per second hitting a sphere of 21.2 cm radius) for different GCR types and solar activity
levels. Values reported here are the integrals of Figure 3.} \label{GCR_rates}
\centering\begin{tabular}{lrr}
\hline
    & Solar min & Solar max \\
GCR Type& rates (\#/s)  & rates (\#/s)  \\
\hline
H      & 6639    & 2452        \\
He     &  649    & 270        \\
Li     &    4.4  &   1.7        \\
C      &   19.7  &   8.2        \\
N      &    5.3  &   2.2        \\
O      &   18.4  &   7.7        \\
\hline
\end{tabular}
\end{table}

\subsection{General Particle Source}

The tool is built in such a way that once geometry and physics models are chosen and set up it can be driven via a scripting
User Interface (i.e. a text macro file) in which the user specifies: (a) the graphical visualization and/or (b)
the type and the number and energy of the incident particles (only part of which will reach the detector).
One single particle with a determined energy represents an event.

We simulated the GCRs seeding via the General Particle Source (GPS; \cite{GSPM-SSD}, \cite{GSPM-Tech}, \cite{GSPM-URD}),
a code for the spacecraft radiation shielding, that is part of the G4 radiation transport code. The user must specify the input
parameters of the particle source: (1) the spatial and (2) the energy distribution. Every time G4 requires a new event, the GPS
will randomly generate it according to the specified distributions. We adopted an angular distribution {\it cosine law} on a spherical
surface (the vacuum sphere), without any preferred particle momentum. The energy distribution follows the SEDs shown in Figure
\ref{GCR_SED} (see also section \ref{checks}).

\section{The Physics Model}\label{PL}
We also had to define a list of physical interactions with the `shield' and the different components of the bolometers. The
model called by our physics list depends on the energy of the incoming particles. The reader may find a detailed list of the
physical processes together with our own choices in \cite{Bon06} and in Appendix B.

\subsection{Thresholds}
In order to avoid the infrared divergence 
some electromagnetic processes require a threshold below which no secondary particles are generated.
Each particle has an associated production threshold, i.e. either a distance, or a range cut-off, which is converted to an energy
for each material, which needs to be defined as an external requirement.
A process can produce the secondaries down to the selected threshold (GEANT4 Users's Guide section 2.4.2).

We have set different production thresholds (hereafter cuts) for each geometrical region. The G4 {\it starting} values, 0.7mm, could
force a too large production of secondary particles for thick volumes such as the `shield', and a too small production for
thin regions such as those of the IPW and the grid. We first ran our G4 code with the {\it starting} cut value of 0.7 mm for all
volumes and then tuned them according to the bolometer geometry. We decided: 1) not to use very small cut values (below 1-2
$\mu$m), since this could affect the validity of physical models at such small steps (Ivantchenko, private communication); 2)
to set cut values $\approx$ 1/5 of the volume thickness. We identified groups of volumes with the same thicknesses: in
particular we defined 5 different regions (see Table \ref{cuts}).
Starting from these assumptions we run simulations with five different cut values per volume, each one varying according to the
thickness of the region under consideration.
\\
We find that the results do not depend strongly on the chosen cuts and the energy depositions
for the different simulations were always consistent with each other to within 1$\sigma$.

\begin{table}
\caption[]{\small The five regions of the bolometer camera and their specific cut values.}\label{cuts}
\centering\begin{tabular}{lrl}
\hline
Region    & Cut Value   & Thickness \\
\hline
Shield    & 2 mm    & 11 mm     \\
Containers& 0.5 mm  & 1.56 - 2.5 mm \\
Si\_Thick & 80 $\mu$m   & 350 - 450 $\mu$m  \\
Si\_Thin  & 15 $\mu$m   & 70 $\mu$m \\
bolometer &  3 $\mu$m   & 0.3 - 5 $\mu$m    \\
\hline
\end{tabular}
\end{table}

\subsection{Theoretical computation of the GCR flux}\label{theorcomp}
In order to disentangle the contributions from the different simulated parts of the instrument (detectors and 'shield') we
evaluated the impact of the GCRs flux on the `naked' detector (without spacecraft shielding). In Table \ref{BmeanHnoSpCnosec}
we report the results from the simulated GCR flux hitting the detector only (results have been normalized to the frame rate
fixed by electronic and software, $1/40\,$s).

We compare them with the theoretical values computed as follows. Let $\Phi$ be the particle flux (in cm$^{-2}$ s$^{-1}$
sr$^{-1}$) and the unit surface $dS$, the number of impacts, $n$ is given by (see i.e. \cite{Sullivan}):

\begin{equation}
n_{impacts} = \Phi \int _{\Omega} \int _{S} \cos \theta d\sigma d\omega = \Phi S 2\pi \int^{\pi/2}_0 \sin\theta \cos\theta d\theta = \pi\Phi S,
\end{equation}
where $S$ is the total superficial area of the volume considered.
This value (in Table \ref{BmeanHnoSpCnosec} the `Th. Hit' column)
should be considered as an upper limit, because it is the maximum theoretical value of the impacts number.

These values (Okumura, 2004, internal report) are then compared with the average hit values obtained from the G4 simulations
for the G4 starting cut values and for our cuts (see Table \ref{BmeanHnoSpCnosec}). We did not take into account for this comparison
the production of secondary events. Although statistics is poor, the results are in good agreement with the
theoretical ones, with a slight better agreement in the latter case where the dispersion around the mean decreases.
Table \label{BmeanHnoSpCnosec} explains the kind of systematic uncertainty introduced by the calculation of the number of impacts per second
as a function of the chosen cuts ("starting cuts" vs "chosen cuts"), i.e. the integration step size in comparison with
geometric part size.

\begin{table}
\caption[]{\small Comparison of the theoretical estimate of the average hit rates (Th. Hit) with the G4 simulations with
'default' cuts and with our choice for them. Calculations were done for the blue bolometer without 'shield' and for two
volumes and solar minimum values.}\label{BmeanHnoSpCnosec}
\centering\begin{tabular}{lccc}
\hline
    &       & Starting Cuts  & Chosen Cuts   \\
Volume  & Th. Hit (/s)  & $\langle n_{impacts}\rangle /s$  & $\langle n_{impacts}\rangle /s$ \\
\hline
Grid    &0.41       &0.5$\pm$0.3          &0.4$\pm$0.1  \\
IPW &2.21       &2.5$\pm$0.9          &2.1$\pm$0.7  \\
\hline
\end{tabular}
\end{table}

\section{Preliminary Tests}\label{checks}

\subsection{The $n_{sec}/n_{prim}$ ratio and production of secondary events}
GCR protons yield a broad variety of secondary products, from photons and electrons to unstable particles (i.e. $\pi^+$), to
some heavier nuclei (i.e. $^{27}$Al). For the other light nuclei, secondary events are mainly photons or electrons. Ion
inelastic scattering may also generate pions, protons and electrons and lighter nuclei, but these events are much rarer.
\\
Secondary events are of utmost importance and are mainly produced in the 'shield'. Their production and their hitting on the
detector chips make them the primary source of `glitches' and largely affect the detector performance.
\\
Secondary events are originated also by the detector chips. As expected the blue chips produce more secondary events, since
they are more massive than the red ones. But the difference between the two arrays is within the error-bars and then the main
secondary events producer is still the 'shield'.

Table \ref{BRnuclei} reports the number of secondary particles produced on average per primary event. We have listed all the
secondaries produced without any energy threshold and independently of their energy release.\\
The $n_{sec}/n_{prim}$ ratio is higher in case of solar max, since, even if GCRs are fewer, they are more energetic, that is
they produce more secondary events.
\\
The $n_{sec}/n_{prim}$
ratio is clearly increasing with the atomic number of the primary particle and the blue and red bolometers have same
dependences, as the two cameras have the same thickness.
Such a ratio is strictly dependent on the physics list we choose and, basically, on the goodness of the G4
models. G4 has been tested for C and lighter nuclei; therefore the ratios for N and O must be then taken with caution. The
minimum proton energy required to produce secondary events is 110eV. 

\begin{table*}
\caption[]{\small The $n_{sec}/n_{prim}$ ratio from simulations of GCR impacts on the bolometers. First column: GCR type;
columns \#2 and \#3: average number of secondary events generated by a single primary event for the blue bolometer in the
cases of solar min and max irradiation; columns \#4 and \#5: same as in columns \#2 and \#3, but for the red
bolometer.}\label{BRnuclei}
\centering\begin{tabular}{lcccc}
\hline
    & \multicolumn{2}{c}{Blue}  & \multicolumn{2}{c}{Red}           \\
    & Solar min & Solar max & Solar min & Solar max         \\
GCR     & $\langle n_{sec}\rangle/n_{prim}$ & $\langle
n_{sec}\rangle/n_{prim}$
    & $\langle n_{sec}\rangle/n_{prim}$ & $\langle n_{sec}\rangle/n_{prim}$     \\
\hline
H       & 2.7$\pm$0.1   & 3.6$\pm$0.1     & 2.6$\pm$0.1   & 3.5$\pm$0.3\\
He      & 6.6$\pm$0.2   & 8.3$\pm$0.1     & 6.6$\pm$0.2   & 8.1$\pm$0.4\\
Li      &11.1$\pm$0.4   &13.5$\pm$0.4     &11.0$\pm$0.2   &12.7$\pm$0.5\\
C       &25.3$\pm$0.4   &33.6$\pm$0.7     &24.6$\pm$0.2   &33.0$\pm$0.4\\
N       &32.3$\pm$0.5   &42.1$\pm$0.9     &31.7$\pm$0.7   &42.1$\pm$0.4\\
O       &38.2$\pm$1.1   &51.6$\pm$0.5     &38.2$\pm$1.1   &51.5$\pm$0.5\\
\hline
\end{tabular}
\end{table*}

\section{Results}\label{results}
The bolometer arrays operate at 300 mK, the electronic readout system has a characteristic frame rate of $1/40\,$s. According
to Table~\ref{GCR_rates} we expect, in this time interval, about 166 proton impacts for a Solar minimum rate and about 61 at
Solar maximum, and very low numbers for heavier ions. To get statistically significant results we should then carry out
thousands of simulations, which would mean a too long computing time. We decided then to increase the number of particles and
then normalize the outputs. We ran, then, five simulations for each nucleus, each one with 2$\times$10$^5$ particles. Results
were normalized to the frame rate of the bolometer, dictated by software and electronics. This choice is further validated (see
below) by the fact that the thermal effects are not cumulative (see below).


\subsection{Deposited energy}
We are particularly interested in the amount of energy deposited on the detector and in the number of pixels affected by one
single GCR. The deposited energy is transformed into heat. The volumetric heat capacity of Si is
c$_{\rm Si}=5.8\times 10^{-7} T^3$ J/K/cm$^3$. At 0.3 K it becomes c$_{\rm Si}=1.57\times10^{-8}$ J/K/cm$^3$.

The enhancement of temperature in the material is then related to the amount of deposited energy as follows:
\begin{equation}
\Delta E (J)= V \cdot c_{\rm Si} \cdot \Delta T (K) = C_{\rm Si} \times \Delta T (K)
\end{equation}

This equation implicitly implies that the efficiency of energy transfer is 100\%. This is an ideal case and formally the
resulting values must be considered as upper limits. However in view of the
high energy involved in the process these values would not be too far from the real ones.
As thermal effects are not cumulative in G4 we have added the deposited energies and then transformed them into temperature values.\\
Impacts on the grid and on the IPW are identified and used to compute the deposited energy that will enhance the temperature
of the bolometers, initially cooled to 300 mK. Tables \ref{BHeat} and \ref{RHeat} report the mean enhanced temperature values (in mK)
for both (blue and red) detectors for shielded simulations.

\subsection{Blue bolometer}
We consider first the effects on the blue bolometer array of the primary particle impacts on the 'shield' and the detector.
The results are shown in Table \ref{BHeat}. The number of hits are given per frame rate of the bolometer.
\\
A {\it glitch} event occurs when the energy deposited by the GCR exceeds some predefined energy threshold,
which has not yet been defined for operation of the bolometers.
We do not apply here any such criterion as an energy threshold cannot be defined yet.
\\
The numbers reported in Table \ref{BHeat} include the {\it total} number of impacts and therefore must be considered as upper limits to the
{\it glitch} events, which will be defined during the data reduction. The values of the deposited energy corresponds to the average
value per hit and $<n>$ to the number of glitches per unit frame rate.
It must be noticed that the word {\it glitch} may not be appropriate here as it does not correspond to a voltage shift but to
a temperature increase. We keep it in this text but advice the reader to consider its meaning as a temperature increase.

\begin{table*}
\caption[]{\small Mean number of impacts per frame rate and the mean heat deposited per hit on the blue bolometer grid with spacecraft shielding
and secondary events, at solar minimum and maximum.}\label{BHeat}
\centering\begin{tabular}{lcccc}
\hline
        & \multicolumn{4}{c}{Blue Grid}     \\
GCR     & \multicolumn{2}{c}{Solar min}     & \multicolumn{2}{c}{Solar max} \\
        & $\langle n\rangle$            & $\langle dT_{dep}\rangle (mK)$
        & $\langle n\rangle$            & $\langle dT_{dep}\rangle (mK)$    \\
\hline
H           & (3.9$\pm$0.5)$\times$10$^{-1}$  &  14.0$\pm$1.8
            & (1.6$\pm$0.1)$\times$10$^{-1}$      &   5.5$\pm$0.5           \\

He          & (0.8$\pm$0.1)$\times$10$^{-1}$  &  6.3$\pm$0.5
            & (4.3$\pm$0.7)$\times$10$^{-2}$      &   2.8$\pm$0.6           \\

Li          & (1.3$\pm$0.1)$\times$10$^{-3}$  &  0.09$\pm$0.01
            & (4.9$\pm$0.8)$\times$10$^{-4}$  &  0.04$\pm$0.01    \\

C           & (2.2$\pm$0.3)$\times$10$^{-2}$  &   1.6$\pm$0.2
            & (0.6$\pm$0.1)$\times$10$^{-2}$  &   0.52$\pm$0.02   \\

N           & (0.7$\pm$0.1)$\times$10$^{-2}$  &  0.52$\pm$0.05
            & (2.7$\pm$0.3)$\times$10$^{-3}$      &  0.21$\pm$0.03    \\

O           & (3.2$\pm$0.2)$\times$10$^{-2}$      &   2.5$\pm$0.2
            & (1.1$\pm$0.1)$\times$10$^{-2}$      &   0.9$\pm$0.1           \\
\hline
\end{tabular}
\end{table*}

\subsection{Red bolometer}
We remind that the red bolometer is made of 2 chips and that the grid is larger than for the blue one (see \S\ref{volumes}).
No other differences with blue bolometer are present. Here we present the results (see Table \ref{RHeat}) of our simulations
normalized to the characteristic frame rate of the bolometer (cfr. above) in the 'shield' + secondary events configuration.

The number of hits between the two arrays differs by a factor 2.6 which is roughly the ratio between the difference in pixels
number (4) and their volume: 4/1.55.

\begin{table*}
\caption[]{\small Same as Table 5 for the red bolometer grid}\label{RHeat}
\centering\begin{tabular}{lcccc}
\hline
        & \multicolumn{4}{c}{Red Grid}      \\
GCR     & \multicolumn{2}{c}{Solar min}         & \multicolumn{2}{c}{Solar max} \\
            & $\langle n\rangle$            & $\langle dT_{dep}\rangle (mK)$
        & $\langle n\rangle$            & $\langle dT_{dep}\rangle (mK)$     \\
\hline
H           & (1.4$\pm$0.2)$\times$10$^{-1}$      & 2.9$\pm$0.5
            & (0.6$\pm$0.1)$\times$10$^{-1}$      & 1.3$\pm$ 0.3            \\

He          & (3.9$\pm$0.6)$\times$10$^{-2}$      & 1.5$\pm$0.2
            & (1.6$\pm$0.3)$\times$10$^{-2}$      & 0.6$\pm$0.1             \\

Li          & (5.1$\pm$0.4)$\times$10$^{-4}$      & 0.021$\pm$0.001
            & (1.8$\pm$0.1)$\times$10$^{-4}$      & 0.009$\pm$0.001   \\

C           & (0.7$\pm$0.1)$\times$10$^{-2}$    & 0.37$\pm$0.02
            & (2.5$\pm$0.2)$\times$10$^{-3}$      & 0.13$\pm$0.02 \\

N           & (2.7$\pm$0.5)$\times$10$^{-3}$      & 0.13$\pm$0.02
            & (1.0$\pm$0.2)$\times$10$^{-3}$      & 0.07$\pm$0.05 \\

O           & (1.2$\pm$0.2)$\times$10$^{-2}$      & 0.56$\pm$0.05
            & (3.9$\pm$0.4)$\times$10$^{-3}$      & 0.21$\pm$0.02 \\
\hline
\end{tabular}
\end{table*}

\subsection{Glitch length}
A large number of glitches (hitting events) are observed on one or two pixels. The theoretical glitch length, i.e. the number
of nearby pixels affected by an event happening in one pixel, can be estimated approximately by simple geometrical
considerations on the solid angle with respect to the bolometer plane (see Table \ref{ng}).

The distribution of incident angles of the particles hitting the detector is isotropic. Some particles will not hit the IPW
but will go straight onto the pixels within a solid angle, $\Omega_1$. These ones will not affect the nearby pixels and do not
contribute to the glitch length. Other particles will hit one or more IPW according to the incident angle value, i.e. solid
angles $\Omega _2$, $\Omega _3$, ... The incident probability is proportional to the solid angle viewed by each pixel and
containing all the incoming directions of particles hitting the IPW. We may therefore compute the ratio of single pixel to
`two pixels glitches' as $\frac{\Omega _1}{\Omega _2}$.\\  Small corrections to this value are also taken into account:
effective cross section for a critical angle, approximation of the circular pixel instead of a square one. The related
uncertainty is given as a lower and upper limit on this theoretical estimation and is given in Table \ref{ng}, where these
values are compared with the results from the average of five proton simulations without a shield.

\begin{table}
\caption[]{\small Comparison between theoretical glitches length (see text for detail) and G4 data for the blue bolometer
(average from 5 simulations of 9441 protons seed each), without 'shield' and without secondary events. In first column the
length in pixel units of a glitch.}\label{ng}
\centering\begin{tabular}{cccr}
\hline
Length & Th$_{up}$ \% & Th$_{low}$ \% & $\langle$G4$\rangle$ \% \\
\hline
2   & 9.77  & 6.52  &14.26$^{+7.19}_{-7.19}$    \\
3   & 3.88  & 2.47  & 1.43$^{+3.19}_{-1.43}$    \\
4   & 1.80  & 1.13  & - \\
\hline
\end{tabular}
\end{table}

We also checked the number of pixels affected by one single GCR for both bolometer cameras. We computed the
percentage of each GCR type producing glitches affecting nearby pixels and assessed that only the adjacent pixels
are affected 20 and 30\% of the time. We also checked the heat propagation effect once a particle hits a IPW: if
we take the values of the IPW volume (see Appendix A) we find that the maximum heat transfer is
$\frac{\Delta T(K)}{\Delta E(MeV)} = 0.48 \frac{K}{MeV}$ along the x-direction and 0.02$\frac{K}{MeV}$ in the y-direction
(where x-y is the bolometer plane). Both effects may not be negligible and deserves further investigation. 
\\
No statistically significant dependencies on solar activity are found.

\section{Observation simulations}\label{discussion}
To test the effect of glitches we have simulated observations of a `strong' and of a `weak' source with the PACS Scientific
Simulator (Andreani et al., in preparation), a complete software code able to simulate from the input sky/source with a given
flux the entire detection chain up to the output recorded signal.

We ran two simulations, one with a thermal source at 50 K with a spectral energy distribution peaking at 55 $\mu$m and very
weak one at a temperature of 5 K.

In the first case the pixels looking at the sources receive an average signal of 3.7 K per pixel; a H glitch yields, on
average, a temperature increase of 24/3700 i.e. of less than 1\%. In the case of the very faint source, at 5 K, the heating
of each pixel due to the source is of 500 mK and the heating produced by the glitch amounts to 24/500, i.e $\sim$ 5\% . The
signal can be accurately retrieved only with a suitable observing technique, allowing more than one pixel (at least a few) to
look at the source, so that the signal lost because of the glitch is recovered from the nearby uncontaminated pixels looking
at the same source.

\section{Test beam results and comparison with simulations}\label{IPN}

We report in this section the comparison between the Geant4 simulations and irradiation tests run in May 2005 at the IPN
(Institute de Physique Nucleaire) in Orsay, Paris.
We ran a complete set of simulations to reproduce the instrumental setup as it is described in \cite{Horeau04,Horeau05}.
Details about the simulations can be found in \cite{Bon05} and the full description of the instrumental setup with
specimen, cryostat and the proton beam are defered in Appendix C (see also \cite{Horeau04,Horeau05}).

One of the blue chip array was hit by a proton and $\alpha$ particles beams. The number of impacts detected in the array and that
simulated are reported in Table \ref{IPNvsG4} (see also \cite{Horeau05}). The different setups during proton irradiations
are listed in the first three columns; column 4 reports the flux computed with the Rutherford's formula, the fifth the same
value reduced by 70\% because of the filling factor discussed in Appendix C\footnote{ this filling factor
was introduced to take into account the experimental
geometry, the bolometer itself and the grid.}. Column 6 lists the measured particle flux and the last one that resulting from
Geant4 simulations.

\begin{table}
\caption[IPN vs. G4]{Comparison of IPN experimental results with the Geant4 output. Column 4 reports the flux calculated with
the Rutherford scattering formula \cite{Rutherford}; column 5 includes the correction for the shadow effect (i.e. the detector sees only 30\%
of the beam); column 6 gives the measured flux; column 7 the flux estimated by Geant4. 
}.\label{IPNvsG4}
\centering\begin{tabular}{ccrrrcr}
\hline
\hline
Particles   & Flux  & Au    & Rutherford scat. & corrected & measured flux &   Geant4\\
        & (part/s)      & (nm)  & particles/s  & particles/s  & particle sum  & particle sum   \\
\hline
proton      & 76.7       &  100  & $\approx$4.3  &   $\approx$1.3    & $\approx$2.25 & 1.86\\
proton      & 127.7      &  100  & $\approx$7.3  &   $\approx$2.2    & $\approx$3.75 & 3.20\\
\hline
\end{tabular}
\end{table}

The experimental {\it Particle sum} value is slightly higher than that given by G4. The difference can be ascribed to either
1) the cut values used in the simulations for the mylar and Al layer placed in front of the detector which may be too large or 2) an
underestimate of the solid angle for the Rutherford diffusion. A decrease of the cut values would increase the number of
secondary events, whereas a bigger solid angle would increase the primary event rate (proton and alpha seeds). At present it
is hard to decide which are the most appropriate modifications of simulation parameters.

\begin{figure}
\centering\includegraphics[height=16cm, width=14cm]{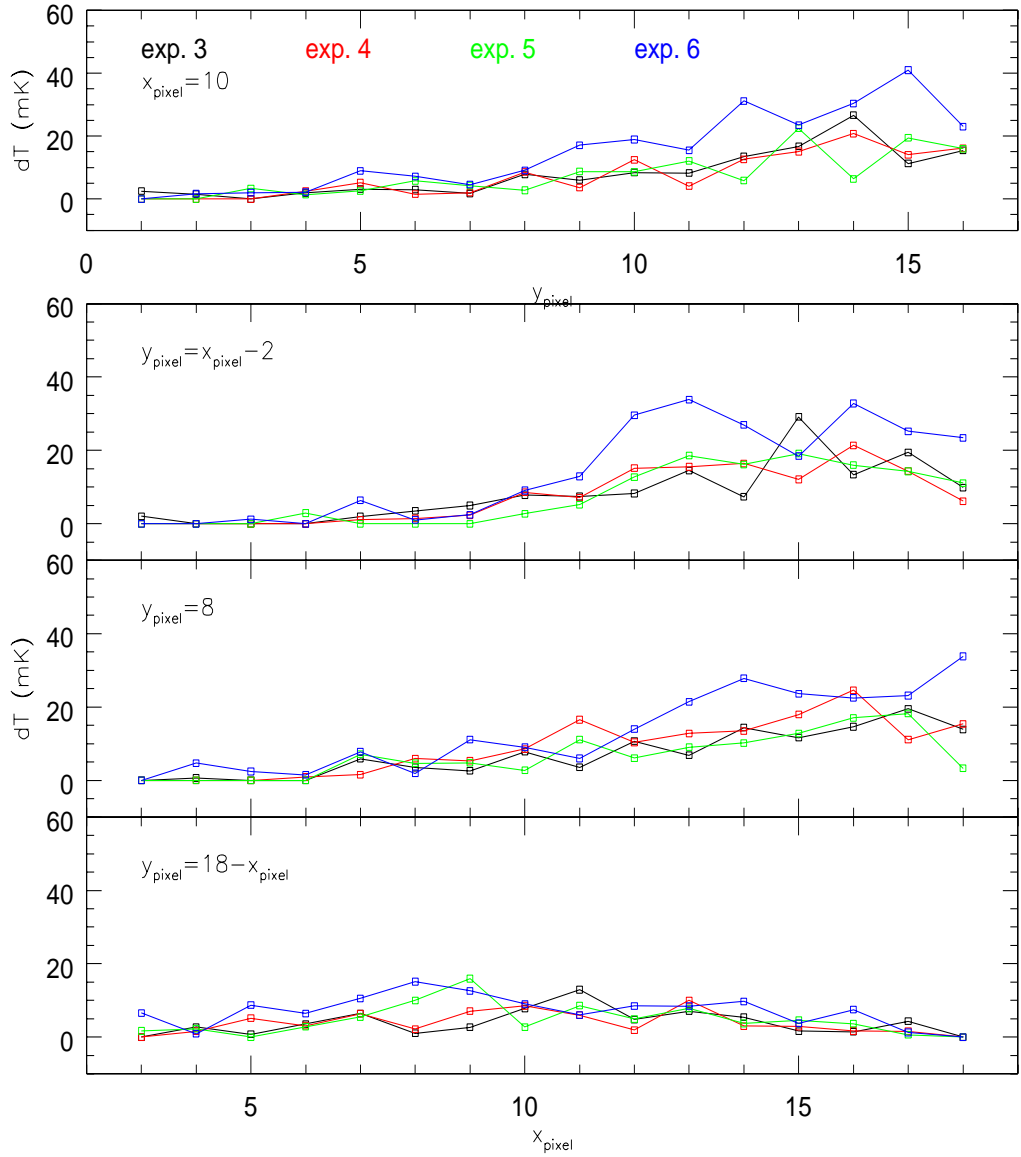}
\caption{Temperature variation per unit time (1s) under proton irradiation along four sections of the $16\times 16$ pixel
matrix, after subtracting the blind pixels: the two median sections (vertical: x$_{pixel}$=10, upper panel; horizontal
y$_{pixel}$=8, second panel from the top) and the two diagonal sections (down-up: y$_{pixel}$=$x_{pixel}$-2, third panel from
the top; up-down: y$_{pixel}$=18-$x_{pixel}$ bottom panel). In black the data referring to the experiment 3 setup, in red for
exp. 4, in green for exp. 5 and in blue for exp. 6. The shadow effect is less present in the plot referring to the up-down
diagonal, whose pixels suffer almost the same shadowing from the cone.}\label{mH}
\end{figure}

Temporal behaviour cannot be easily retrieved from the present simulations and in this case we exploit the shadowing effect
of the beam cone (see Appendix C and Figure \ref{shadoweffect})
to check for temperature variation as a function of the incoming flux.
We have compared the pixels which were only slightly affected
by the incoming beam (maybe even not hit by any particle) with those clearly subject to a strong bombardment. We report in
Figure \ref{mH} the temperature variation along four imaginary cuts performed along the pixel matrix. The temperature varies
by at most 15\% for pixels receiving the largest number of hits. Note that the number of hits on pixel columns 1 to 5 differ
by a factor $\ge 10$ from that on columns 12 to 16. This result was also found in the measurements and is due to the
geometrical effect of the cone.

These results show that the particles incoming on a pixel increase its temperature by up to 15\%. The pixel-to-pixel
variation yielded by our simulations, due to the increasing intensity of the beam, may be understood in terms of temporal
behaviour of a single pixel under the effect of a cosmic particle. This may result in a slow responsivity drift, i.e. in a
change of the response of the single pixel with a time scale depending on the bolometer time constant.

\section{Conclusions: effects on observations}\label{Conclusion}
\begin{itemize}
\item Red bolometers are less affected by GCR glitches than the blue ones. The pixel grid of the red bolometer is a factor of
1.569 larger so that its heat capacity (volumetric heat capacity times the volume) is correspondingly higher for the red
pixels. The number of impacts is proportional to the detector surface. Both the red and the blue bolometer are affected by
the same impinging flux but they have different collecting areas: 2048 grids for the blue and 512 grids for the red and the
expected ratio of hit rate is: $\frac{A_{\rm blue} \cdot 2048}{A_{\rm red} \cdot 512} = \frac{173599.7 \mu ^2 \cdot
2048}{272392.1 \mu ^2 \cdot 512} \sim 2.55$, where $A_{\rm blue}$ and $A_{\rm red}$ are the array areas of the blue and red
bolometers respectively.

\item We have quantified the contribution of secondary events produced by the 'shield'. This latter acts as the major source
of secondary events, while the bolometer itself produces approximately 1/3 of them.

\item The sensitivity of the bolometer cameras may be limited by glitch impacts for very faint sources; we expect a glitch
contamination $\le 5$--6\%  of the background. The contamination is lower for the red camera.

\item A test with the instrument simulator shows that in case of very faint sources a suitable observing technique and glitch
removal algorithm must be put in place. In particular this latter should work on each chopping time interval in order to
avoid false detections due to the sudden change of the flux level.
\end{itemize}

\begin{acknowledgements}
We warmly thank M. Asai, G. Cosmo, A. De Angelis, F. Lei, V. Ivantchenko, G. Santin and J.P. Wellish, who helped us in this
effort and elucidated the tricks of the Geant4 code. We acknowledge the help of many people at CEA (Saclay) who provided us
with the details of the bolometer geometry.
\end{acknowledgements}

\section{Appendix A: The Detector Design}\label{volumes}
In the G4 code a class describing the geometry and chemical composition of the
detectors must be set. Here we describe how the geometrical model for the PACS bolometers is built.

The detector module is made of different parts: a grid, an Inter-Pixel-Wall, different layers (Al, Si,
Ti), a Si strip, reflecting slabs, etc. \cite{bolocea}. Each pixel is supported by a silicon grid
5$\mu$m thick with a thermometer, and is suspended in an Inter-Pixel-Wall (IPW) grid 450$\mu$m thick
which has a reference thermometer and acts as a thermostat. The detector is `decomposed' in different
cells, each of a simple geometrical shape (parallelepipeds (called boxes by G4) and spheres). Each part
is characterized by its shape, its chemical composition and spatial position (in a Cartesian 3D space
coordinates). The output signal is a function of the temperature difference between the pixels and the
reference thermometers.\\
Attached to them two columns of blind pixels heated at a mean temperature of nominal pixels are used to
suppress any undesired electronic temporal deviations. All the components of the detector are at a
temperature of 0.3 K.

\begin{itemize}
\item The four aluminum layers have been approximated as boxes \footnote{In case of a box shape we gave
its measures as x$\times$y$\times$z.}. One large layer (43$\times$50.88$\times$2.5 mm) placed at the
central spatial coordinate system (0,0,0). A second small Al layer (37.95$\times$50.88$\times$2.5 mm)
lying horizontally on the large layer. Two more layers, vertically placed and tangentially to the x base
of the two previous layers (43$\times$1.56$\times$9.7 mm).
\item The bolometer camera is made of two different arrays: the blue one consists of 4$\times$2 chips
and the red one of 2$\times$1 chips. Each chip contains the following components (single or multiple):
\begin{enumerate}
\item One Titanium Layer : an L-shaped volume which is simulated with two boxes lying one under the
other. The large box (21.7$\times$12.2$\times$2 mm) is placed 0.4 mm above the small Al layer. Every Ti
box is detached by the other by 0.55 mm. The smaller Ti box (we called it 'Ti edge',
3$\times$12.2$\times$1.9 mm) is lying below and on the external parts of the large Ti layer.
\item The Silicon Layer (Si Layer) is a unique box (19.15$\times$12.63$\times$0.45 mm) placed 0.8 mm
above the large Ti layer. Each Si layer is detached one by the other by 0.12 mm.
\item the 16$\times$18 pixels of each chip have three identical Reflecting Slabs made of Titanium,
Palladium and Gold. They lie one on top of each other over the Si layer and have dimensions of
640$\times$640$\times$0.3 $\mu$m
\item Each pixel has a Silicon grid which is approximated by a continuum slab with a volume equivalent
to the real one and a thickness equal to the real one. It turns out that the slab has a projected (x,y)
area smaller that the effective grid. We draw it as a box of dimensions 416.653$\times$416.653$\times$5
$\mu$m, for the blue bolometer, and (521.912$\times$521.912$\times$5$\mu$m), for the red bolometer.
The volumes of the absorbers are then 868000 $\mu$m $^3$ for the blue and 1361960 $\mu$m $^3$ for the red.
\\
Analysis of the different covering factor of the slab, compared to the real grid will be given below.
\item 17$\times$19 Indium spheres (radius 10 $\mu$m) are lying on top of the Si layer. They are placed
at the intersection of the Inter-Pixel-Wall (IPW), sustaining it.
\item The Inter-Pixel-Wall made of Si defines the border of the Si grid. The IPW, pole-on viewed, is a
grid and it was decomposed in similar parts: 5 different boxes are built making the whole IPW. 2+2 boxes
make the border of the IPW thicker respect to the internal IPW (the ones separating the grids): (1) the
longest boxes (horizontal border, HBWall) are 14.13$\times$0.35$\times$0.45 mm; (2) the shortest boxes
(vertical border, VBWall) are 0.35$\times$11.93$\times$0.45 mm.(3) The external vertical border (Extra
Wall) is thicker than the usual VBWall. That is we added an extra box of 1.65$\times$12.13$\times$0.45
mm.\\ The IPW grid was designed by considering (4) 15 long (horizontal, HWall) boxes of
13.41$\times$0.07$\times$0.45 mm each, with (5) 17 short boxes (vertical, VWall) placed vertically to
the long ones (0.07$\times$0.68$\times$0.45 mm each).
\item The Mask (blind pixels, 2$\times$16).\\
Both the bolometers have two rows of blind pixels. They are equal to the other pixels, but front masked
by different strata of different materials. These blind pixels are placed in the external part of the
chip. There are three slabs of the same material and in the same order of the reflecting layers, that is
Ti, Pd and Au, designed as boxes 3.125mm$\times$12.63mm$\times$0.3$\mu$m each. Above them there is a
thick Silicon 3.125$\times$12.63$\times$0.45 mm layer.
\end{enumerate}
\end{itemize}

\section{Appendix B: The Physics Model}\label{AppendixB}
We define the list of physical interactions with the shielding and the different components of the
photoconductors in order to best reproduce the impact of GCRs on the photoconductors. The list is
comprehensive of all the possible physical interactions:

\begin{itemize}
\item Electro-magnetic physics: photo electric and Compton effect; pair production; electron and
positron multiple scattering, ionization, bremsstrahlung and synchrotron; positron annihilation.
\item General physics: decay processes.
\item Hadron physics: neutron and proton elastic, fission, capture and inelastic processes; photon,
electron and positron nuclear processes; $\pi^+$, $\pi^-$, K$^+$, K$^-$, proton, anti-proton,
$\Sigma^+$, $\Sigma^-$, anti-$\Sigma^+$, anti-$\Sigma^-$, $\Xi^+$, $\Xi^-$, anti-$\Xi^+$,
anti-$\Xi^-$, $\Omega^+$, $\Omega^-$, anti-$\Omega^+$ and anti-$\Omega^-$ multiple scattering and
ionization. We added the High Precision Neutron dataset, that is valid for neutrons till 19.9 MeV.
\item Ion physics: multiple scattering, elastic process, ionization and low energy inelastic processes.
We added inelastic scattering (J.P. Wellish, private communication) using the binary light ion reaction
model and the Shen cross section \cite{Shen}.
\item Muon physics: $\mu^+$, $\mu^-$ multiple scattering, ionization, bremsstrahlung and pair
production; $\mu^-$ capture at  rest; $\tau^+$ and $\tau^-$ multiple scattering and ionization.
\end{itemize}

For 'ion physics' of $alpha$-particles, protons, and H isotope nuclei we used the cross section by Tripathi
et al. (\cite{Tripathi}), holding for energies of up to 20 GeV per nucleon. The cross sections used for the other ions
\cite{Ivan03} are valid only up to 10 GeV per nucleon. This means that we had to cut out high energy ions of
the GCRs (whose flux is lower than 10$^{-5}\,$/m$^2$/s/sr). The treatment of ion physics is strictly valid
only for nuclei not heavier than C. However we have applied it also to simulate N and O, which should not
behave much differently.  

The model called by our physics list is different depending on the energy of the incoming particles. Our
physics list is based on QGSP\_BIC\footnote{Quark Gluon String Physics\_Binary Cascade} list (J.P. Wellish,
private communication; see the Geant4 web page User Guide \S 2.4.3 for details). Modifications to such a list are:

\begin{enumerate}
\item We added the High Precision Neutron dataset, which is valid for neutrons up to 19.9 MeV.
\item For neutrons and protons, we introduced\footnote{It is well known that (1) the BIC
does not reproduce well data for energies below 50 MeV \cite{Ivan03}. In contrast, the Bertini
cascade works well below 50 MeV for all but the lightest nuclei. And (2) the BIC for pions and kaons
has not been tested yet (J.P. Wellish, private communication).} the Bertini cascade \cite{Bertini} up to 90
MeV and made the BIC starting from 80 MeV.
\item QGSP is assumed valid from 10, rather than 12, GeV.
\item We have taken into account the gamma- and electro-nuclear reactions, so we considered the
electromagnetic\_GN physics list.
\end{enumerate}

\section{Appendix C: the description of the set up of the irradiation measurements}
The detector used for the proposed investigations under proton and alpha particle irradiation was the
detector module SMD 3 blue (N$^{\circ}$
05-05 type MV), i.e. the Flight Model (equivalent doping) mounted on a common carrier in flight configuration. The ``3''
indicates the position on the common carrier. This means that the mounting is as if the bolometer was complete, so 7
locations out of 8 on the mount were empty. Both the bolometer and the reaction chamber were placed in vacuum. Two pumps
held the vacuum to $\approx 10^{-6}$ mbar.
\\
The proton energy in the experimental setup was 20 MeV and the flux could be changed from 76.6 to 127.7
particles per second.\\
The beam, before reaching the blue bolometer elements penetrates several layers and undergoes Rutherford diffusion due to a
45$^{\circ}$ inclined gold foil placed at 1m from the detector, decreasing the flux of the beam. The foil has two different
thicknesses according to the running test. Proton tests were all run with a 100 nm thick foil.

\begin{figure}
\centerline{\includegraphics[height=12cm, width=12cm]{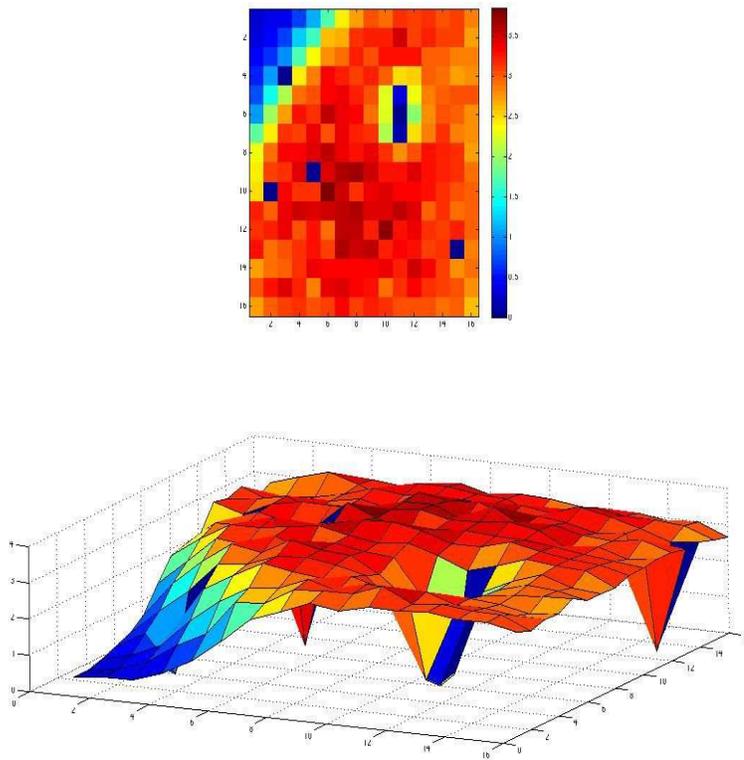}}
\caption[]{\small The picture shows the number of observed glitches/s/pixel. On the left the shadow due to the cone is
visible. Considering the surface of the bolometer arrays (16*16 pixels or bolometers) only 30\% of it is "filling"
(due to geometrical configuration, see text for details).
}\label{shadoweffect}
\end{figure}

The diffused beam then encounters a couple of 14$\mu$m thick Al foils, placed 75 cm from the detector. Then it crosses a
45$^{\circ}$ inclined foil of mylar (a polymer: [-O-C::O-C6H6-C::O-O-CH2-CH2-]n) 25 $\mu$m thick; it goes through a sort of
cave cone.
Then the beam hits the blue bolometer module. Because of the detector geometry -- the bolometer itself does not consist of a
full Si chip but of a grid -- the surface exposed to the cone is not 100 percent full. In addition the cone itself introduces
a shadow effect on the bolometer because the cone face-on area masks part of it. This can be seen in Figure
\ref{shadoweffect} where the number of glitches per second and pixel are shown in a three dimensional plot. A shadow effect
is present on the left, the bolometer array receives less hits because of the cone. Considering the surface of the bolometer
array only 30\% ``sees'' the beam.

\end{document}